# Magnetic field dependence of kinetic inductance in $Bi_2Sr_2Ca_2Cu_3O_{10}$ superconducting strip and its feasible applications

S. Sarangi[*], S. P. Chockalingam, S. V. Bhat

Department of Physics, Indian Institute of Science, Bangalore-560012, India

[*]Corresponding author:

    Subhasis Sarangi

    Department of Physics

    Indian Institute of Science

    Bangalore – 560012, India

    Tel.: +91-80-22932727, Fax: +91-80-23602602

    E-mail: subhasis@physics.iisc.ernet.in

**Abstract:**

The kinetic inductance properties of a thin superconducting sample of BSCCO-2223 compound is studied. A strong dependence of kinetic inductance with the variation of applied magnetic field is found at low temperature in the superconducting state. The kinetic inductance of a thin superconductor increases linearly with increasing magnetic field. This behavior is used for designing a superconducting modulator circuit. This behavior can be useful for device applications and designing of superconducting electronic circuits like superconducting power amplifier, superconducting transmitter and superconducting ac current controller.

**Introduction:**

Resistors, capacitors and inductors are known as the linear passive devices in the electronic circuits. Inductors are used in the electronic circuits to control ac current without energy loss unlikely resistors, which lead to $I^2R$ loss always. Power factor of a pure inductive circuit is always 0 [1]. The presence of finite resistance in the copper based inductors restricts them for some of the important electronic applications but in the other hand, inductors made off superconductors can be very useful for various electronic applications due to zero resistance and zero power factor. A straightforward technique of controlling the inductance of a superconducting strip by varying the applied magnetic field is demonstrated. A superconducting modulator circuit is built by using this technique.

In the case of metallic inductors, the total inductance comes only from the magnetic part known as magnetic inductance $L_m$. But the inductance associated with a superconductor $L_s$ consists of two parts: the magnetic inductance $L_m$, defined by the geometry of the sample, and the kinetic inductance $L_k$, defined by the inertia of the superconducting carriers. Magnetic inductance caused by the magnetic energy stored due to the magnetic field in the dielectric medium outside the inductor and the kinetic inductance caused by the energy stored due to the kinetic energy of superelectrons inside the superconductor. Kinetic inductance is related to the magnetic penetration depth of the superconductors [2]. For example, because of the large magnetic penetration depth of the high $T_c$ superconductors, the kinetic inductances of the high $T_c$ superconductors are larger than the low $T_c$ superconductors. Therefore, the high $T_c$ superconductors are the ideal choice for making devices working in the principles of kinetic inductance. The magnetic field dependences of the kinetic inductance of various superconductors are studied to find out its common origin and the possibilities of device making. The standard four-probe ac impedance measurement technique is used for studying the kinetic inductance at different magnetic fields, frequencies and temperatures.

The kinetic inductance has strong temperature dependence just below $T_c$ and is used for photon detection, imaging and making coplanar wave guide, thermometer and microwave components [3, 4, 5, 6, 7, 8, 9, 10]. Most of the experimental studies of the kinetic inductance have been limited so far by the temperature range near the superconducting transition. This is advantageous for device operation since the number of quasiparticle is exponentially small and the corresponding generation-recombination noise is small as well in this range of temperature [11]. So most of the devices suggested in the literature work just below $T_c$. But there are certain demerits of working just below the critical temperature or in the transition region itself. The first demerit is that the devices working in or near the transition region always accommodate with some finite value of resistance and the second demerit is that a small disturbance can drag the device to the resistive region. So it is always advantageous and safe to work far below the transition temperature. The results discussed in this paper are observed well below the superconducting transition and can be useful for device making.

The magnetic field and the thickness dependence of the kinetic inductance is studied in details well below $T_c$. Application of magnetic field has no effect on the magnetic inductance coming from the geometry of the sample. But it is possible to change the kinetic inductance of a superconducting strip by applying magnetic field. It is also possible to change the kinetic inductance by varying the strip thickness. So a desired value of kinetic inductance can be achieved from a superconducting strip either by changing the strip thickness or sweeping the applied magnetic field. These results have important implications for the understanding of the kinetic inductance behaviors in a thin superconductors at low temperature and useful for designing superconducting electronic circuits.

**Experiment:**

In these experiments, the high $T_c$ cuprate superconductor BSCCO-2223 samples are used. The BSCCO powder used was made by a conventional solid-state reaction method. High purity powders of $Bi_2O_3$, $SrCO_3$, $PbO$, $CaCO_3$, $CuO$ and $Ag_2O$ were mixed in stoichiometric proportion, calcinated at $830^0$ C for 24 hour and sintered at $860^0$ C in air for 170 h. The calcinations and grinding procedure were repeated three times. The sample shows sharp superconducting transition temperature at ~111 $K$ as determined by ρ~$T$ and ac susceptibility measurements. The materials were found to be single phased as determined by x-ray diffraction.

The kinetic inductance of the superconducting strips is measured in the low frequency range upto 100 kHz by measuring the imaginary part of the complex ac impedance. In the complex ac impedance measurement, the in-phase signal (real part) shows the resistivity and the out-phase signal (imaginary part) shows the reactive part of the sample. Special care has been taken to minimize the capacitive contribution. Rectangular strips with dimensions of *11 × 4 × 3, 11 × 4 × 2, 11 × 4 × 1* mm$^3$ were cut from the sample pellets for the four probe complex ac impedance measurements. Four terminal complex ac impedance measurements were performed with a standard dual phase lock-in-amplifier (model SR830) with two-phase detections. The gaps between the conjugative terminals in the 4-probe measurement were 2 mm (see the inset of Fig. 1 for the schematic of the experimental arrangements). The current and voltage leads were soldered onto the surface of the samples with silver paint contacts. The contact resistances were estimated to be below 2 Ω. The ac current was monitored across a standard resistor in series with the sample, and the sample voltage was measured using the lock-in-amplifier. The sample was arranged inside an Oxford instrument cryostat. AC impedance studies have been performed from 10 to 300 K. The ac impedance was measured in the frequency range 100 Hz < $f$ < 100 kHz. The dispersive behaviors of the leads are carefully checked by using some standard cells (metallic samples with very low resistance) and ensured that there was no extraneous inductive or capacitive coupling in this frequencies range. Magnetic field was varied upto 1.2 T using a Bruker electromagnet. The magnetic field was scanned slowly with the rate of 100 Gauss per minute to minimize the electric signals coming from the rapid flux induction. The oxford cryostat was mounted in between the two poles of the electromagnet. In these experiments, the contributions from the magnetic inductance $L_m$ and the lead inductance $L_{lead}$ were measured and found to be very less compared to the kinetic inductance, so the impedance values described in the figures are considered to be only due to the kinetic inductance.

**Results & Discussion:**

Figure 1 shows the temperature dependence of the imaginary parts of the ac impedance of the BSCCO strip of dimension *11 × 4 × 2 mm* at different magnetic fields (1000 G, 500 G and 0 G) measured at frequency of 1 kHz. The imaginary part in this figure is considered to be only due to the kinetic inductance after neglecting other sources of inductances. The superconducting transitions are clearly visible at the temperature of 111 K. The imaginary part of the ac impedance at low temperature increases with increasing magnetic field but the change in the imaginary part of the ac impedance with magnetic field reduces with increasing temperature towards the critical temperature. Below the transition temperature, the imaginary part of the ac impedance is a function of penetration depth ($\lambda$). Inset in figure 1 shows the real parts of the ac impedance of the same BSCCO strip. For all the three magnetic fields, the real parts of the ac impedance remain alike. We have assumed that both the resistive and the inductive responses coexist in the resistive state of a certain part of the superconducting transition.

Figure 2 shows the variation of both the real and imaginary parts of the ac impedance of the BSCCO strip with the magnetic field measured at the frequency of 1 kHz and the temperature of 20 K. The strip dimensions is *11 × 4 × 2* mm$^3$. The real part shows the resistance of the sample and it is independent of magnetic field up to 1000 Gauss. The imaginary part $\omega L_k$, which is due to the kinetic inductance, varies linearly with the magnetic field up to 1000 Gauss. The linear dependence of the kinetic inductance with the magnetic field does not hold at very high magnetic field. The imaginary part ($\omega L_k$) increases with magnetic field in the range of field from 0 to 1000 Gauss with an average change of 80 × $10^{-9}$ µΩ / Gauss. Inset in Fig. 2 shows the block diagram of the ac impedance measurement unit.

Figure 3 shows the frequency dependence of the real and imaginary parts of the ac impedance in the frequency range from 100 Hz to 100 kHz at zero magnetic field and 20 K temperature. The impedance due to the kinetic inductance ($\omega L_k$) increases with an average change of 20 × $10^{-9}$ µΩ / Hz and the in-phase part of the ac impedance (resistance) is nearly independent of frequency. Inset in Figure 3 shows the relation between the kinetic inductance and the thickness of the superconducting strip (*d*) at 20 K temperature and 1 kHz frequency. The kinetic inductance ($L_k$) increases with decreasing the thickness of the strip. The kinetic inductance ($L_k$) is inversely proportional to the thickness of the superconducting strip

So from the above results it can be concluded that it is possible to vary the inductive reactant of a superconducting strip within a broad range by varying either the thickness of the strip or the applied magnetic field. Now it is important to understand the effect of applied magnetic field, strip thickness and ac frequency on the imaginary part of the ac impedance before we start discussing about the possible applications of these results and designing of the superconducting modulator circuit. The application of magnetic field or the variation of temperature alters the London penetration depth (λ) of superconductors, which in turn changes the kinetic inductance associated with it. For a superconducting sample, imaginary part of the ac impedance equals $L_k \varepsilon \omega$, where $L_k$ is the kinetic inductance caused by the non-dissipative motion of the superconducting electrons and ε is a complex frequency dependent dielectric constant quantity important only in the presence of vortices. The contribution from kinetic inductance comes into the picture when we pass ac current. In the case of ac current, an electric field must be present to accelerate the electrons. Because of the inertial mass of the electrons (even if of very small magnitude), the supercurrent lags behind the electric field. Hence the superelectrons contribute inductive impedance known as the kinetic inductance $L_k$. The kinetic inductance $L_k$, defined by the inertia of the superconducting carriers is expressed as

$$L_K = \left(\frac{m_q}{nq^2}\right)\left(\frac{l}{\sigma}\right) \qquad \text{-----------------------(Equation 1)}$$

For a uniform current in a homogeneous conductor, the kinetic inductance $L_k$ depends on the length $l$ and the cross-sectional area $\sigma$ of the conductor. It also depends on the material through the number density $n$ of the current carriers, their mass $m_q$, and their charge $q$ [2].

The rise of the imaginary part of the ac impedance with the magnetic field is due to the increase of the kinetic inductance, which is due to the result of superconducting pairing. Unlike normal conductor, in a superconductor, electric current flows only in the penetration depth ($\lambda$) layer (a well-known simple solution of the London equation and a consequence of the Meissner effect) in the absent of external magnetic field. In the presence of a magnetic field one has to think about effective penetration depth (Campbell length) and more complex frequency - dependent response [12, 13]. The total penetration depth in the mixed state is $\lambda^2 = \lambda_L^2 + \lambda^2_{vortex}$ where $\lambda_L$ is the London penetration depth and $\lambda_{vortex}$ the contribution from vortex motion. A compressive expression for $\lambda_{vortex}$ has been derived by several authors [14, 15, 16]. At low temperature and frequencies well below the pinning frequencies (of order GHz in cuprates), $\lambda_{vortex}$ reduces to the Campbell pinning penetration depth

$$\lambda_C^2 = \frac{\phi_0 B}{4\pi\alpha} \quad\text{-----------------------(Equation 2)}$$

Here $\phi_0$ is the flux quantum and $\alpha$ is the Labusch parameter [12, 17]. From the above equations it is clear that the Campbell length increases with magnetic field. It is well established that the kinetic inductance $L_k$ of a superconducting strip of length $l$, width $w$, and thickness $d$ can be expressed in a two fluid model as

$$L_K = \mu_0 \lambda^2 \frac{l}{wd} \quad\text{-----------------------(Equation 3)}$$

Combining the above two equations, a linear field dependence of Campbell length is obtained as

$$\omega L_K = \omega \mu_0 \left(\frac{\phi_0 B}{4\pi\alpha}\right)\frac{l}{wd} = \left(\frac{\omega\mu_0\phi_0 l}{4\pi\alpha\omega d}\right)B = C_k B \quad\text{--------------------(Equation 4)}$$

$C_k$ is the proportionality constant which value is 80 × 10$^{-9}$ μΩ / Gauss for the superconducting strip at the temperature of 20 K. The imaginary component of the ac impedance is mainly due to the kinetic inductance so it increases with increasing magnetic field (Fig. 1). In the other hand the inductive reactance is directly proportional to the applied ac frequency so the imaginary component of the ac impedance of the superconducting strip increases linearly with frequency (Figure 2). Kinetic inductance is inversely proportional to the area of cross-section σ (see Equation 1). This explains the thickness dependent behavior in the inset of Figure 2.

**Modulator Circuit Design and Implementation:**

A superconducting modulator circuit is designed by using these results. A superconducting modulator circuit works like a signal multiplier in an electronic circuit. Two ac signals $V_1$ and $V_2$ get multiplied or modulated. The schematic circuit diagram of a superconducting modulator circuit is illustrated in Figure 4. The basic superconducting modulator circuit consists of a capacitor $C$ and the superconducting strip connected in series. The strip is surrounded by a magnetic head, which produces magnetic field proportional to the supply current. So the supply current to the magnetic head controls the inductance of the superconducting strip. Low frequency signal is given to the magnetic head as the base supply and the high frequency signal is given to the superconducting strip through the capacitor as the main supply. In this configuration the circuit works as a superconducting modulator, the output ac voltage $V_{out}$ can be expressed as the modulation of the ac signals $V_1$ and $V_2$. Some of the basic functions of superconducting modulator circuit are implemented by modulating two separate signals of frequencies 100 kHz and 1 kHz by using the above design. We modulate the 1 kHz signal by oscillating the external applied magnetic field at 1 kHz. We use C = 10 µF as a best match for modulating the 1 kHz signal with the 100 kHz signal.

**Conclusions:**

In conclusion, a strong correlation between the kinetic inductance and the applied magnetic field is observed in the superconducting strip of BSCCO. It is interpreted that the phenomenon is due to the increasing of Campbell length with the increasing magnetic field. This phenomenon is used for designing a superconducting modulator circuit. The discussed phenomenon is also observed in other superconducting strips made of high $T_c$ superconductors like YBCO, LSCO and MgB$_2$. This phenomenon can be useful for device applications and designing other superconducting electronic circuits.

**Acknowledgements:**

This work is supported by the Department of Science and Technology, University Grants Commission and the Council of Scientific and Industrial Research, Government of India.

**Figure Captions:**

1. Temperature dependence of the imaginary part (mainly due to the kinetic inductance) of the complex ac impedance of the superconducting strip of dimensions *11 × 4 × 2 mm* at 1 kHz ac frequency and different magnetic fields of 1000 G, 500 G and 0 G. The inset shows the temperature dependence of the real part (due to the resistance) of the complex ac impedance of the superconducting strip at 1 kHz ac frequency and different magnetic fields of 1000 G, 500 G and 0 G. There is no significant variation in the real part is observed by changing the magnetic field.

2. Magnetic field dependence of the real (due to the resistance) and imaginary (mainly due to the kinetic inductance) parts of the complex ac impedance of the superconducting strip of dimensions *11 × 4 × 2 mm* at 1 kHz ac frequency and at 20 K temperature. The inset shows the block diagram of the 4-probe ac impedance measurement circuit. The input ac current to the sample comes from the signal output of the lock-in amplifier and the output ac voltage from the sample goes to the signal input of the lock-in amplifier for detection.

3. Frequency dependence of the real (mainly due to the resistance) and imaginary (mainly due to the kinetic inductance) parts of the complex ac impedance for the superconducting strip of dimensions *11 × 4 × 2 mm* at 20 K temperature and zero magnetic field. The inset shows the kinetic inductance behavior as a function of the thickness of the superconducting strip of dimensions *11 × 4 × 1 mm*, *11 × 4 × 2 mm* and *11 × 4 × 3 mm*.

4. Circuit diagram of a signal modulator. The signal $V_1$ is modulated with the signal $V_2$. The output $V_{out}$ is the multiplication of both the input signals $V_1$ and $V_2$. $L$ is the total inductance of the superconducting strip. The superconducting strip is surrounded by the magnetic head. The magnetic field is oscillated at the frequency of the ac signal $V_1$.

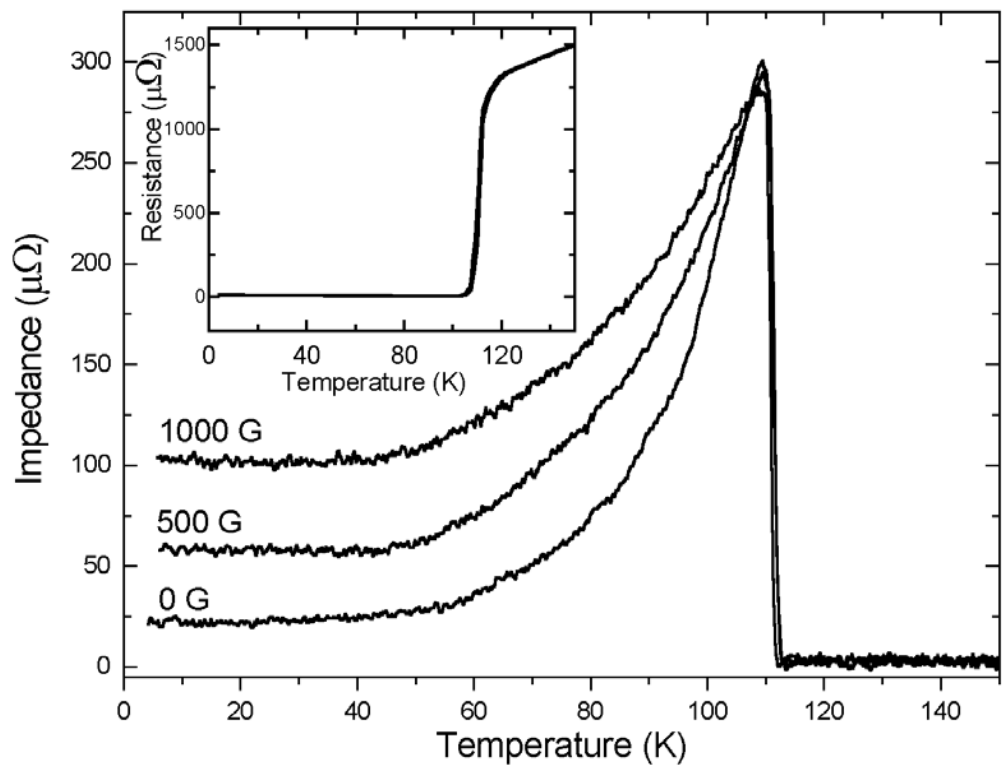

**FIG. 1.**

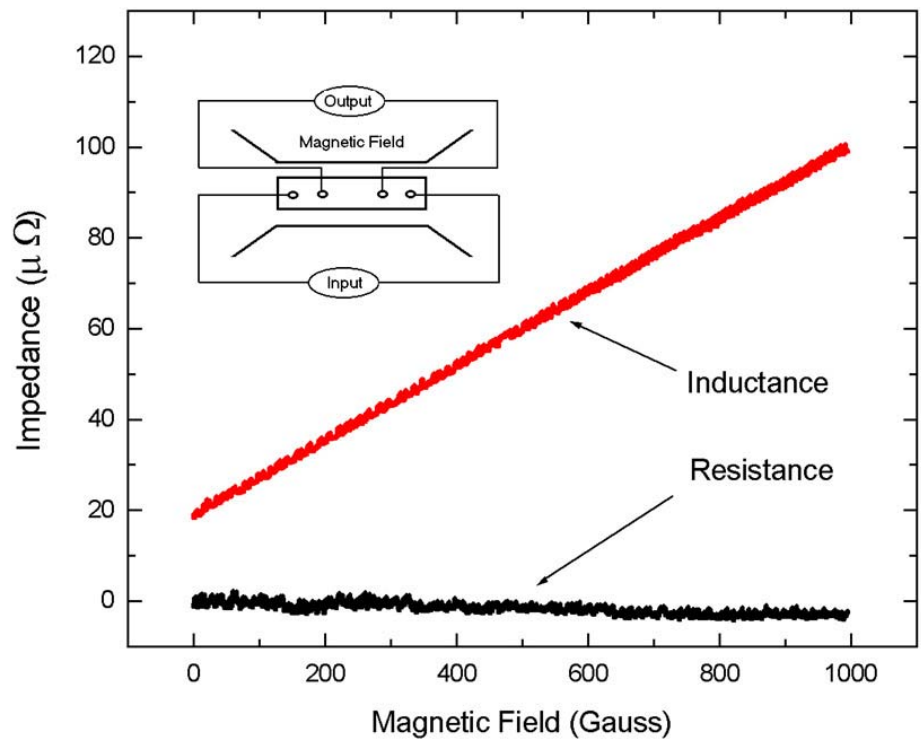

**FIG. 2.**

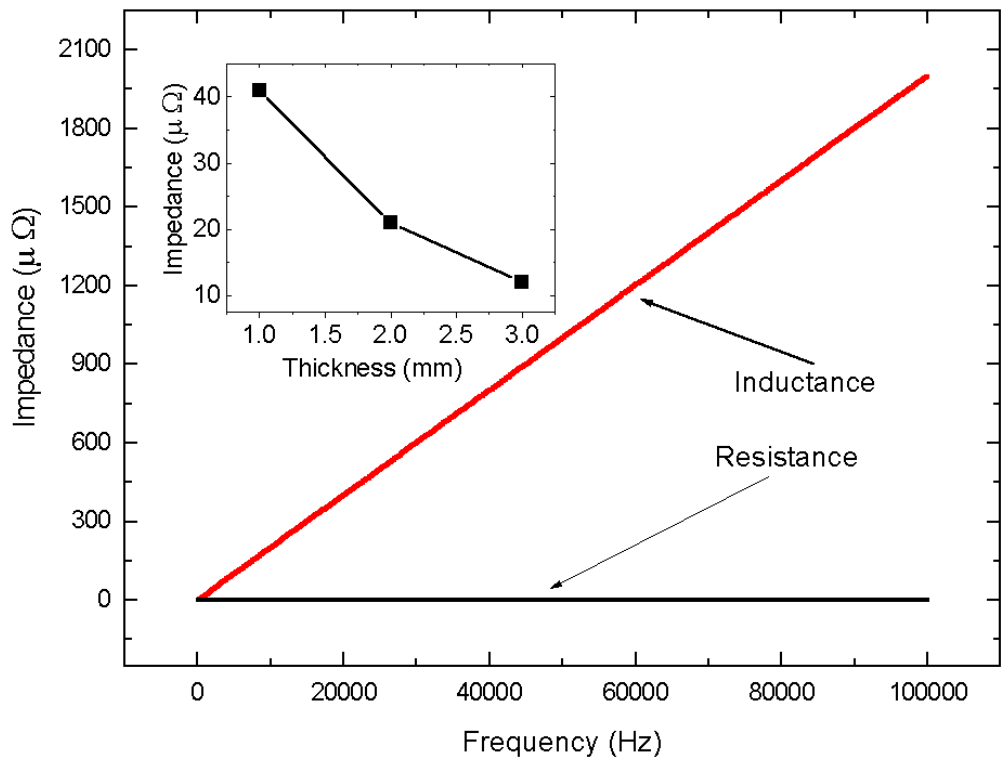

**FIG. 3.**

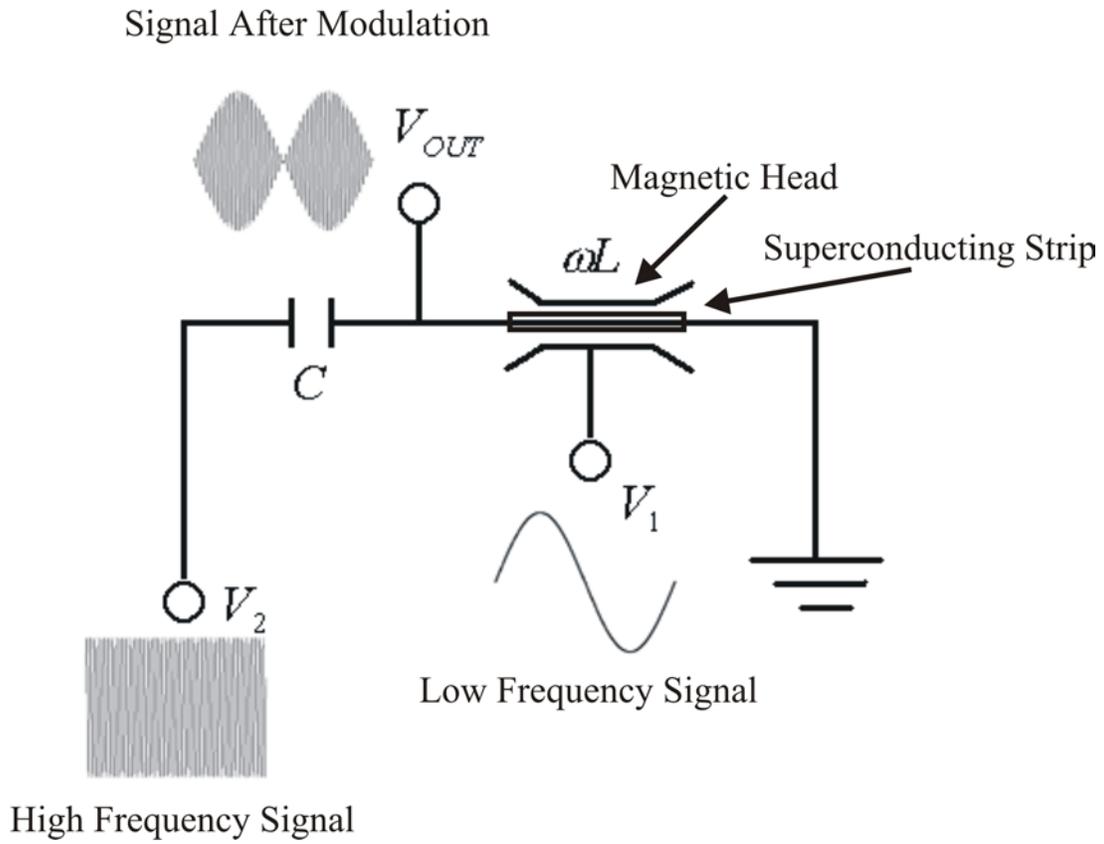

**FIG. 4.**